\documentclass[10pt,letterpaper]{article}
\usepackage{opex3}
\usepackage{epsfig}
\usepackage{cite}
\usepackage{color}
\newcommand{\D}{\Delta}
\newcommand{\p}{\ensuremath{\hat p}}
\newcommand{\q}{\ensuremath{\hat q}}
\renewcommand{\a}{\ensuremath{\hat a}}
\newcommand{\be}{\begin{equation}}
\newcommand{\ee}{\end{equation}}
\newcommand{\ba}{\begin{eqnarray}}
\newcommand{\ea}{\end{eqnarray}}
\newcommand{\ave}[1]{\ensuremath{\langle #1 \rangle}}
\def\lsim{\mathrel{\rlap{\lower4pt\hbox{\hskip0pt$\sim$}}
    \raise1pt\hbox{$<$}}}                
\def\gsim{\mathrel{\rlap{\lower4pt\hbox{\hskip0pt$\sim$}}
    \raise1pt\hbox{$>$}}}

\begin{document}


\title{The quest for three-color entanglement: experimental investigation 
of new multipartite quantum correlations}

\author{K. N. Cassemiro,$^1$ A. S. Villar,$^{1,2}$ M. Martinelli,$^1$ and P. Nussenzveig$^{1 *}$}

\address{$^1$Instituto de F\'\i sica, Universidade de S\~ao Paulo,
Caixa Postal 66318, \\ S\~ao Paulo, SP, Brazil, 05315-970 \\
$^2$Current address:  Institute for Quantum Optics and Quantum Information of the \\ Austrian 
Academy of Sciences, and Institut f\"ur Experimentalphysik, \\ Universit\"at Innsbruck, Technikerstr. 25/4, 6020 Innsbruck, Austria \\ 
$^*$Corresponding author: \textcolor{blue}{\underline{nussen@if.usp.br}}} 



\begin{abstract} 
We experimentally investigate quadrature correlations between pump, signal, 
and idler fields in an above-threshold optical parametric oscillator. We 
observe new quantum correlations among the pump and signal or idler beams, as 
well as among the pump and a combined quadrature of signal and idler beams. 
A further investigation of unforeseen classical noise observed in this system 
is presented, which hinders the observation of the recently predicted 
tripartite entanglement. In spite of this noise, current results approach the 
limit required to demonstrate three-color entanglement. 
\end{abstract}



\section{Introduction}

The optical parametric oscillator (OPO) is a very well-known source of non-classical light. Quantum correlations between 
the intensities of the downconverted beams, signal and idler, were first measured already in 1987~\cite{firsttwin87}. Because 
of the high degree of correlation attainable between these two bright light fields they were called, in the above-threshold 
operation, \textit{twin beams}. In fact, squeezing in the intensity difference reached the impressive value of 
$-9.7$~dB~\cite{fabre10dB}. As was first recognized by Reid and Drummond in 1988, the quantum correlations are not 
restricted to intensity but also extend to their phases: the OPO produces twin beams in an {\em Einstein-Podolsky-Rosen} 
(EPR) state~\cite{reiddrummondepropo_prl88}. EPR-type entanglement was obtained a few years later in the OPO, but only 
when operating below threshold~\cite{kimbleepr}. Since then, this system has been used in many applications in 
continuous-variable quantum information~\cite{furusawatele_science98,pengdensecoding_prl02,furusawanetwork_nature04}. 

The measurement of entanglement in the above-threshold OPO would take 17 years to be 
realized~\cite{prlentangtwinopo, optlettpeng, pfisterentang}, in part because of technical difficulties to perform 
phase measurements on bright beams that usually have different frequencies. Another important reason was the presence 
of an unexpected excess noise generated in the intracavity pump field by the non-linear crystal, which was transferred 
to the sum of downconverted phases during the parametric process~\cite{josaboptquinfo}. As a consequence, squeezing 
in this observable could only be found very close to the oscillation threshold~\cite{prlentangtwinopo}. This strange 
feature, unaccounted for by the theoretical model, went unnoticed until recently, when phase measurements were 
performed above threshold. We present here a further study of the fields' quadrature noises as functions of pump power to 
show that the standard theoretical model for the OPO can not account for the observed features. Even the 
{\em ad hoc} model presented in ref.~\cite{josaboptquinfo} reveals itself to be insufficient. These noise 
features warrant investigation in themselves, which will have to go beyond the present work.

Another ``surprise'' was to be encountered in this thoroughly studied system. Although quadrature squeezing 
of the reflected pump beam was experimentally observed in 1997~\cite{kasaisqzpump_europhyslett97}, little 
attention was paid to quantum properties of the pump field. In many theoretical approaches, the pump was 
treated as a classical quantity. A very recent investigation of the quantum properties of the full three-mode system, 
by {\em Villar et al.} revealed that all three fields should be entangled~\cite{prltrientangopo}. This seems 
to be the simplest system predicted to directly generate genuine tripartite continuous-variable entanglement. 
It also appears as a distinctive feature of above-threshold operation, in contrast to the widely-used 
OPO below threshold. The first observation of a triple quantum correlation was recently reported by 
{\it Cassemiro et al.}, who observed squeezing in the combination of the sum of twins' phases with the reflected 
pump amplitude, for a detuned OPO~\cite{optletttripqucorr}. This result was a good indication that phase--phase 
correlations should be present for a resonant OPO, as theoretically predicted~\cite{prltrientangopo}.

Here we present the first direct observation of such triple quantum phase correlations. For higher pump powers, 
we show as well that the amplitude of the reflected pump beam is quantum-mechanically correlated to each of the twin's 
amplitudes. As for the tripartite entanglement, owing to the spurious excess phase noise introduced by the 
non-linear crystal in the intracavity pump field~\cite{josaboptquinfo}, the quantum correlations observed do 
not suffice. 

The quest for bright multicolor entangled beams will be rewarded by providing an interesting tool for 
quantum information networks, since they can concomitantly interact with physical systems otherwise incompatible. 
We describe below our latest experimental advancements in this direction.

\section{Tripartite entanglement}

The existence of direct tripartite pump-signal-idler entanglement can be physically understood as a consequence of energy conservation. 
The parametric process imposes for the frequencies of these fields the relation $\nu_0$=$\nu_1$+$\nu_2$ (indices 0, 1, and 2 
refer to pump, signal, and idler fields, respectively). Phase fluctuations, regarded 
as small frequency fluctuations, are therefore connected: $\delta\varphi_0$=$\delta\varphi_1$+$\delta\varphi_2$. This leads to strong 
quantum correlations among the phase fluctuations of the three fields. On the other hand, the creation of a pair of signal and 
idler photons inside the OPO cavity only occurs upon annihilation of a pump photon. In triply resonant cavities, as we consider 
here, pump depletion above threshold is a well known effect. Correlations between the amplitudes of the pump and twin beams are 
thus created.

In order to demonstrate genuine multipartite entanglement, a more rigorous approach makes use of inequalities which must be 
fulfilled by all separable states. A violation of a subset of the inequalities suffices. We denote the electromagnetic 
field's amplitude and phase quadratures as $\p_j=\exp(-i\varphi_j)\a_j$ $+\exp(i\varphi_j)\a_j^{\dag}$ and 
$\q_j=-i[\exp(-i\varphi_j)\a_j-\exp(i\varphi_j)\a_j^{\dag}]$, respectively, where $j\in\{0,1,2\}$. The fields' annihilation and 
creation operators are denoted $\a_j$ and $\a_j^\dag$, satisfying $[\a_j,\a_{j'}^\dag]=\delta_{j,j'}$, $j'\in\{0,1,2\}$, and 
$\varphi_j$ are the fields' mean phases. Tripartite entanglement among three fields can be witnessed by violation of the following 
inequalities~\cite{prltrientangopo, vanloockmultipartite_pra03}:
\begin{eqnarray}
\label{crit1}
V_0&=&\D^2\left(\frac{\p_1-\p_2}{\sqrt{2}}\right)+ \D^2\left(\frac{\q_1+\q_2}{\sqrt{2}}-\alpha_0 \q_0\right) \geq 2 ,\\
\label{crit2}
V_1&=&\D^2\left(\frac{\p_0+\p_1}{\sqrt{2}}\right)+ \D^2\left(\frac{\q_1 -\q_0}{\sqrt{2}} +\alpha_2 \q_2\right) \geq 2 ,\\
\label{crit3}
V_2&=&\D^2\left(\frac{\p_0+\p_2}{\sqrt{2}}\right)+ \D^2\left( \frac{\q_2-\q_0}{\sqrt{2}} +\alpha_1 \q_1\right) \geq 2 ,
\end{eqnarray}
where $\alpha_j$ are numbers that minimize $V_j$. These are generalizations for tripartite systems of a criterion for 
bipartite systems derived independently by Duan {\it et al.} and Simon~\cite{dgcz_prl00,simon_prl00}. In fact, if one 
makes $\alpha_j=0$, the bipartite criterion is recovered. Violation of at least two inequalities entails genuine 
tripartite entanglement. In our case, one has $V_1=V_2$ according to the model, since signal and idler beams play the 
same role in the OPO equations, differing only by polarization for type-II conversion, and by optical frequency. In the experiment, 
these two inequalities differ slightly owing to imperfections, such as unbalanced losses for signal and idler intracavity 
modes. 

The first inequality contains well-known correlations. Its first term, $\Delta^2\p_-$ [where $\p_-=(\p_1-\p_2)/\sqrt{2}$] 
corresponds to the amplitude correlation between the twin beams. The second one, $\Delta^2\q'_+$ [where 
$\q'_+=(\q_1+\q_2)/\sqrt{2}-\alpha_0\q_0$], can be written as the difference of $\Delta^2\q_+$ [where 
$\q_+=(\q_1+\q_2)/\sqrt{2}$], which involves the twins' phase anti-correlations, and a correction term 
$\beta_0$ given by their phase correlation with pump,
\be
\label{ineqtriplecorr}
\Delta^2\q'_+=\Delta^2\q_+-\beta_0\;, \qquad \beta_0= \frac{(C_{\q_0\q_1}+C_{\q_0\q_2})^2}{2 \,\D^2\q_0},
\ee
where $C_{\q_j\q_{j'}}= \ave{\delta\q_j \delta\q_{j'}}$, $j'\neq j$. Notice that $\beta_0\ge0$. If 
$\Delta^2\q'_+ <0$ and $\beta_0\neq0$, there is a quantum correlation between all three fields. The inequality corresponding 
to $V_0$ is violated even in the absence of the correction term, since signal and idler are entangled~\cite{prlentangtwinopo}. 
The correlation to the pump, signaled by $\beta_0$, strengthens this entanglement and increases the region of parameters 
where violation occurs~\cite{optletttripqucorr}.

Here, we present an extended experimental investigation of this system, by measuring in addition the quadrature noises 
appearing in the second and third inequalities. To simplify notation, we named the terms appearing in $V_1$ and $V_2$ as
\ba
\D^2\p_{0j}&=&\D^2\left(\frac{\p_0+\p_j}{\sqrt{2}}\right),\\
\D^2\q'_{0j}&=&\D^2\left(\frac{\q_j -\q_0}{\sqrt{2}} +\alpha_j \q_j\right).
\ea
Following Eq.~(\ref{ineqtriplecorr}), we write
\be
\label{ineq23}
\Delta^2\q'_{0j}=\Delta^2\q_{0j}-\beta_{j'}, \qquad \beta_{j'}= \frac{(C_{\q_0\q_{j'}}-C_{\q_j\q_{j'}})^2}{2 \,\D^2\q_{j'}},
\ee
where $\q_{0j}=(\q_j-\q_0)/\sqrt{2}$ and $j'\neq j$. In this case, $\q_{0j}$ is the composite noise of the pump and 
one of the twin beams, and $\beta_{j'}$ is a correction term directly dependent on their correlations with the remaining beam. 

\section{Experimental set-up and system characteristics}

The optical parametric oscillator comprises a non-linear $\chi^{(2)}$ crystal disposed inside a linear Fabry-Perot cavity. 
The coupling mirror for the pump beam is a partial reflector at 532~nm (reflectivity $R=$69.4\%) and a high reflector 
($R>$99.8\%) at 1064~nm; the coupling mirror for the twin beams has reflectivities which are equal to 96.0\% at 1064~nm 
and greater than 99.8\% at 532~nm. The crystal is a 12~mm long type-II High Gray Tracking Resistant Potassium Titanyl 
Phosphate (KTP) from Raicol Crystals Ltd. The crystal temperature is kept constant by a peltier element near 25~$^\circ$C, 
with a stability of the order of tens of mK. Its losses are approximately 3\% at 532nm and less than 1\% at 1064nm. 
Typical threshold power is about 75~mW, and the OPO cavity bandwidth is $\delta \nu_{opo}$=45(2)~MHz. A frequency doubled 
diode-pumped Nd:YAG source (Innolight Diabolo) at 532~nm is used to pump the OPO. This laser is `filtered' by a 
mode-cleaning cavity~\cite{modecleancavity} (bandwidth of 2.4~MHz) prior to injection in the OPO, in order to remove 
all classical noise which could hinder the entanglement for analysis frequencies above 15~MHz~\cite{josaboptquinfo,optcomm04}. 

Phase fluctuations are measured through a self-homodyne technique using optical cavities~\cite{galatola_optcomm91, optcomm04}. 
An optical cavity adds a frequency-dependent phase to the field it reflects, as a function of the detuning between the incident 
beam carrier frequency and the cavity resonance frequency. Scanning the cavity resonance (e.g. with a piezo-electric device to 
change its length) results in different dephasings between carrier and sidebands for each cavity detuning. In the Fresnel 
plane this can be qualitatively visualized as a rotation of the noise ellipse with respect to the mean field. As a result, 
the intensity fluctuations of the beam reflected by the cavity can correspond to fluctuations of any desired quadrature 
of the incident beam. It can be shown that the incident phase quadrature is totally converted into the reflected amplitude 
quadrature when the analysis frequency $\Omega$ is higher than $\sqrt{2}$ times the cavity bandwidth 
$\delta\nu$~\cite{galatola_optcomm91, optcomm04}. The cavity-carrier detunings $\Delta$ (relative to the cavity bandwidth) 
for which it occurs are $\Delta\approx\pm0.5$ and $\Delta\approx\pm\Omega/\delta\nu$.

\begin{figure}[ht]
\centering
\epsfig{file=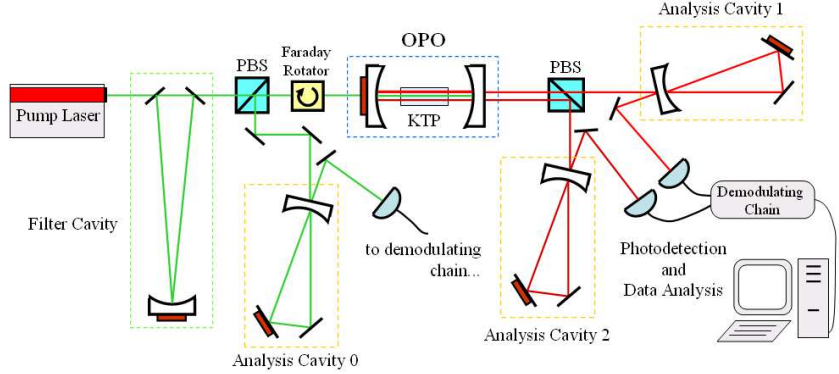,scale=0.45}
\caption{Sketch of the experimental setup. PBS: polarizing beam splitter.}
\label{tripasetup}
\end{figure}

A sketch of the experimental set-up is presented in Fig.~\ref{tripasetup}. There are three analysis cavities in the 
apparatus, with the following bandwidths:  $\delta \nu_0$=11.5(6)~MHz, $\delta \nu_1$=14.5(1)~MHz and $\delta \nu_2$=13.6(2)~MHz, 
where the indices correspond to the beam analyzed in each one. In order to measure the pump beam reflected by the OPO we 
use a Faraday Rotator and a polarizing beam splitter (PBS). The infrared beams are detected with high quantum efficiency 
(93\%) photodiodes (Epitaxx ETX300), resulting in an overall detection efficiency of $\eta=87(3)$\%. Another photodetector 
is used for the pump beam (Hamamatsu S5973-02, quantum efficiency of 94\%), with an overall detection efficiency of $\eta_0=74(3)$\%.

The photocurrent high frequency component (HF) is sent to a demodulation chain, where it is mixed with a sinusoidal reference 
at the desired analysis frequency ($\Omega=$~21~MHz), within a 600~kHz bandwidth. Analysis frequency is chosen to have the smallest 
value (since the quantum correlations increase for decreasing frequency) such that the analysis cavity can still completely 
rotate the beams' noise ellipses. The resulting low frequency beat note is acquired at a rate of 600~kHz by an analog-to-digital 
(A/D) card connected to a computer. Variances of each individual noise component are then calculated taking a group of 1000 
acquisition points. Finally, the noise power is normalized to the standard quantum limit (SQL), previously 
calibrated~\cite{optletttripqucorr}.

\section{Measurement results}

Our measurements were performed with the OPO cavity closely resonant to all three fields. We estimate that the OPO cavity 
detunings were smaller than 0.05 relative to its bandwidth for twin beams. In this situation, amplitude and phase quadratures 
decouple~\cite{optcomm04}. All analysis cavities were synchronously scanned, such that the same quadrature (amplitude, phase, 
or a linear combination of them) was acquired simultaneously for the three beams at each time. 

\begin{figure}[ht]
\centering
\epsfig{file=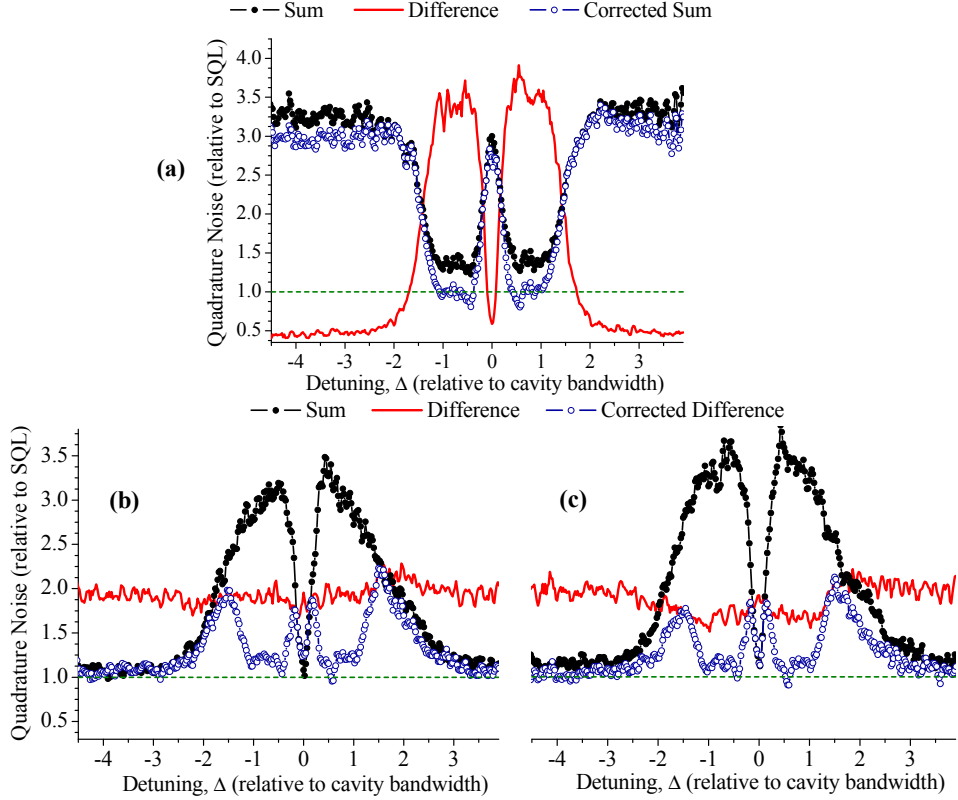,scale=0.75}
\caption{Noise measurements of the terms appearing in the tripartite entanglement criteria of Ineqs.~(\ref{crit1})--(\ref{crit3}) 
(pump power relative to threshold $\sigma=P/P_{th}=1.14$). As functions of cavity detuning $\Delta$, black full circles + line 
curves correspond to the noises in the sum of quadratures of two beams, solid red line to their difference, and blue open 
circles + line to a corrected noise, in which information from the remaining beam is included. Amplitude quadratures are 
measured at large cavity detuning $|\Delta|\gsim 3$ and at $\Delta=0$; phase quadratures are measured at $\Delta=\pm0.5$. (a) The 
considered pair of beams is signal and idler, and the corrected {\em sum} curve includes the term $\beta_0$ coming from 
correlations with the pump beam; (b) signal--pump beams and {\em difference} correction using idler; (c) idler--pump beams 
and {\em difference} correction using signal. The corrected sum in (a) presents the first direct observation of a quantum 
correlation among the phases of pump, signal, and idler: the noise is 26\% below the SQL.}
\label{tripa-cav}
\end{figure}

In our experimental investigation towards tripartite entanglement, we obtain data as presented in Fig.~\ref{tripa-cav}. It 
shows the noise of the sum (full circles + line) and the difference (full line) of each pair of beams' quadrature fluctuations 
as functions of the analysis cavities' detunings $\Delta$ relative to their bandwidths (set to be equal during the experiment). 
The calibrated shot noise level is given by the dashed line. Fig.~\ref{tripa-cav}(a) shows signal-idler combined noises, 
Fig.~\ref{tripa-cav}(b) considers signal-pump beams, and Fig.~\ref{tripa-cav}(c) depicts idler-pump noises. All curves show 
the expected noise ellipse rotation: amplitude quadrature noises are measured for large detunings ($|\Delta|\gsim 3$) and exactly 
on resonance ($\Delta=0$), while phase quadrature noises appear for detunings half-bandwidth off-resonance ($\Delta\approx\pm0.5$). 

A third curve in each figure (open circles + line) shows the effect of including information about the third beam. They are 
obtained by subtracting the correction term from the sum curve, in the case of Fig.~\ref{tripa-cav}(a), or from the difference curve, 
in the cases of Figs.~\ref{tripa-cav}(b)-(c). The relevant regions in the corrected curves are located at $\Delta\approx\pm0.5$, 
corresponding to phase measurements [(a) $\Delta^2\q'_+$, (b) $\Delta^2\q'_{01}$ and (c) $\Delta^2\q'_{02}$]. All the curves 
presented in Fig.~\ref{tripa-cav} come from a single measurement where all three fields' noises were recorded at the same time.

We begin our analysis by Fig.~\ref{tripa-cav}(a), devoted to signal-idler correlations and to the noise correction by pump beam. The 
well-known squeezing in $\p_-$ is seen for large detunings in the difference curve, with noise value $\Delta^2\p_-=0.45(1)$. The sum curve 
at $\Delta=\pm0.5$ gives the noise in $\q_+$, equal to $\Delta^2\q_+=1.28(2)$. Although no squeezing is observed in $\q_+$ in this situation 
(pump power relative to threshold $\sigma=P/P_{th}=1.14$), entanglement is present, since $\Delta^2\p_-+\Delta^2\q_+=1.73(3)<2$, 
violating the Duan {\it et al.} inequality~\cite{dgcz_prl00}. However, squeezing can be obtained in the three-field composite 
observable $\q'_{+}$, with the value $\Delta^2\q'_{+}=0.84(2)$, according to the corrected sum curve at $\Delta=\pm0.5$. This is the 
first direct observation of a quantum correlation among the phases of pump, signal, and idler. Ineq.~(\ref{crit1}) then provides 
the expected violation of the first tripartite entanglement inequality, $V_0=\Delta^2\p_-+\Delta^2\q'_+=1.29(2)<2$.

Figs.~\ref{tripa-cav}(b) and (c) present pump-twin correlations, and their difference noise corrected by the remaining twin. 
These two are very similar, as expected. The first term appearing in Ineq.~(\ref{crit2}) is measured in the sum curve of 
Fig.~\ref{tripa-cav}(b). It is shot noise limited, $\Delta^2\p_{01}=1.03(3)$, as expected by the theoretical model, for this 
pump power. The difference of their phases $\Delta^2\q_{01}$ presents excess noise, which is reduced by the 
correction term $\beta_2$. The corrected difference phase noise $\q'_{01}$, however, is not squeezed as expected, but shot 
noise limited as well, $\Delta^2\q'_{01}=1.01(4)$. This saturates Ineq.~(\ref{crit2}), $V_1=\Delta^2\p_{01}+\Delta^2\q'_{01}=
2.04(5)\approx2$. A similar result holds for Fig.~\ref{tripa-cav}(c). In this case, $\p_{02}$ presents a small amount of 
excess noise, $\Delta^2\p_{02}=1.12(2)$, and $\q'_{02}$ is shot noise limited, $\Delta^2\q'_{01}=0.97(3)$, resulting in 
$V_2=\Delta^2\p_{02}+\Delta^2\q'_{02}=2.09(4)>2$ for Ineq.~(\ref{crit3}).

\subsection{Unexpected noise features}

\begin{figure}[ht]
\centering
\epsfig{file=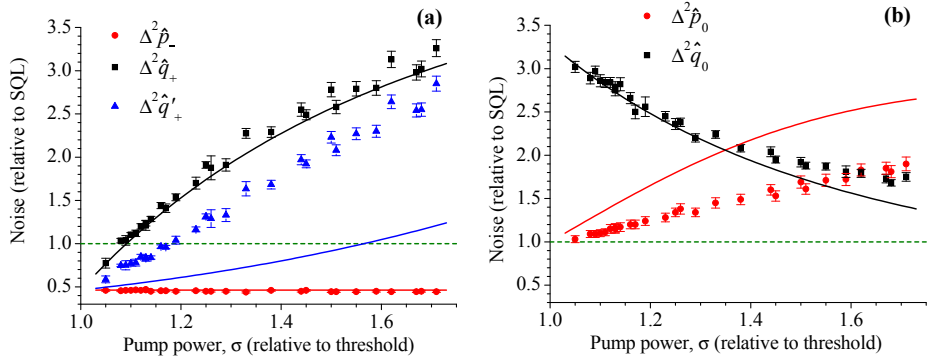,scale=0.65}
\caption{Noise behaviors as functions of pump power $\sigma$. (a) Terms appearing in Ineq~(\ref{crit1}): the well-known 
squeezing in the subtraction of twins' intensity fluctuations $\Delta^2\p_-$ is shown in the red circles, noise of twins' sum 
of phase fluctuations $\Delta^2\q_+$ are the black squares and the phase noise corrected by information from the pump 
$\Delta^2\q'_+$ are the blue triangles. (b) Amplitude (red circles) and phase (black squares) noise of the pump beam reflected 
by the OPO cavity. Solid lines are tentative fits to the {\it ad hoc} model explained in the text. Predictions differ from 
the usual ones only for the phase quadratures. Theoretical curves have the same colors as the experimental curve to which 
they relate.}
\label{behavsigma}
\end{figure}

Contrary to our expectations~\cite{prltrientangopo}, we could not observe squeezing in $\Delta^2\q'_{01}$ 
or in $\Delta^2\q'_{02}$. We performed several measurements similar to those presented in Fig.~\ref{tripa-cav} and recorded 
the values of noises for various pump powers. Each set of curves taken in one measurement scan, as the ones presented in 
Fig.~\ref{tripa-cav}, gives seven quantities of interest: $\Delta^2\p_-$, $\Delta^2\q_+$, $\Delta^2\p_0$, $\Delta^2\q_0$, 
$\Delta^2\q'_+$, $\Delta^2\p_{01}$, and $\Delta^2\q'_{01}$. Part of the results are presented in Fig.~\ref{behavsigma}, 
where quadrature noises or their combinations are plotted as functions of pump power relative to threshold, $\sigma$.

In Fig.~\ref{behavsigma}(a), the behaviors of terms belonging to Ineq.~(\ref{crit1}) are presented. The only noise curve 
that behaves in agreement with the theoretical model is $\Delta^2\p_-$ (red circles). It's constant value 
is quantitatively compatible with what we expect from the other independently measured OPO parameters (mirror transmissions, 
total intracavity loss, OPO bandwidth, and detection efficiency~\cite{optcomm04}), $\Delta^2\p_-=0.46(1)$. Model 
prediction is the red solid line. 

Noise in the sum of twins' phases $\Delta^2\q_+$ (black squares) shows squeezing only very close to the oscillation threshold 
($\sigma\leq1.1$) and excess noise otherwise. The usual theoretical model predicts that it should be squeezed 
for all values of $\sigma$ tested, reaching shot noise only for $\sigma\gg 1$, if the incident pump beam is shot noise limited. 
The pump beam at the output of the filter cavity was verified to fulfill this requirement~\cite{josaboptquinfo}. Yet, 
as we reported previously, excess noise is introduced in the intracavity pump beam by interaction within the non-linear 
crystal~\cite{josaboptquinfo,optletttripqucorr}. This was observed in the pump beam reflected from the OPO cavity, even 
when no parametric oscillation took place. The data presented in those papers was successfully modeled by introducing 
extra noise in the input pump beam. As we speculated in~\cite{josaboptquinfo}, this could be due to a non-linear 
refractive index effect, which would generate excess phase noise. The quasi-resonant cavity would then partially convert it into 
amplitude noise. In order to circumvent this effect, we decided to lower the pump finesse, which would minimize 
accumulated spurious phase shifts (the authors of ref.~\cite{optlettpeng} claim this to be the main difference between 
their system, in which they do not observe this extra phase noise, and others). The OPO cavity finesse for the pump beam 
here is much lower than in our previous experiments. This modification entailed a lower reflected pump excess phase noise and 
completely suppressed the reflected pump amplitude noise but, alas, the mystery is far from solved, as described below.

In Fig.~\ref{behavsigma}(b), we present measurements of the reflected pump amplitude and phase quadratures during oscillation, 
respectively shown as red circles and black squares, as functions of $\sigma$. We could not account simultaneously for 
the data of Figs.~\ref{behavsigma} (a) and (b). We tried to mimic the spurious excess noise effect by introducing {\it incident} 
pump phase noise $\Delta^2\q_0^{in}$ in the equations, as an {\it ad hoc} free parameter. Two different values of this 
parameter are needed in order to obtain the (already poor) agreement observed in Figs.~\ref{behavsigma} (a) and (b). In 
order to obtain the black solid curve in Fig.~\ref{behavsigma} (a), we used the value $\Delta\q_{0T}^{in}=14$; in order 
to do the same in Fig.~\ref{behavsigma} (b), we used $\Delta\q_{0P}^{in}=6$. The reflected pump phase noise is lower than 
would be needed to account for the twins' excess phase noise. Furthermore, the reflected pump amplitude noise is even 
lower than expected by the linear model with {\em no} extra noise (already taking into account the finite detection 
efficiency). 

Next, we observe a correlation between the pump and twins' phase quadratures. In spite of its novel manifest quantum 
nature, we do not find agreement with the model. The blue solid line in Fig.~\ref{behavsigma} (a), was obtained by 
using $\Delta^2\q_{0T}^{in}=14$. Since we are investigating a correlation between pump and twins' 
quadratures, and we could not find a single parameter to account for pump and twins separately, the poor 
agreement is not surprising.  

\begin{figure}[h!t]
\centering
\epsfig{file=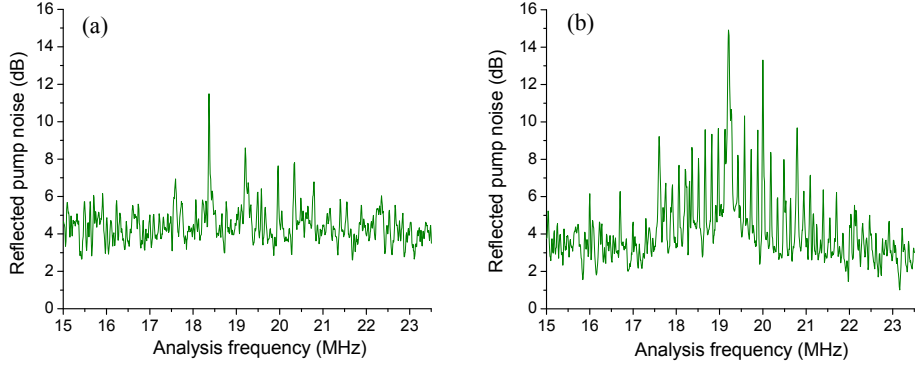,scale=0.6}
\caption{Reflected pump phase noise, as a function of the analysis frequency. Pump power is close to 
threshold, but temperature is tuned to avoid triple resonance. In (a), the pump polarization is such 
that the phase matching conditions are fulfilled, whereas in (b) the orthogonal polarization is used. 
Measurements are normalized to the shot noise level (0 dB level).}
\label{peakspump}
\end{figure}

The noise features of the reflected pump phase quadrature are indeed more complicated than could be 
modeled by a single parameter. We measured this phase noise as a function of the analysis frequency, for 
pump powers close to threshold and two different incident polarizations.In this measurement, the analysis cavity 
for the pump beam is locked to a detuning $\Delta=0.5$ and the photocurrents are analyzed by means of a 
spectrum analyzer (Resolution Bandwidth, RBW = 10~kHz, Video Bandwidth, VBW = 100~Hz). The phase matching 
conditions for the parametric process are only fulfilled for one of these polarizations. We observed a rich 
spectrum, presented in Fig.~\ref{peakspump}, with evenly spaced peaks [spacing approximately 150(10)~kHz] of up 
to 15~dB of excess noise, on top of a broad $\sim 4$~dB excess noise plateau. The peak structure is more pronounced 
for the ``wrong'' polarization, for which phase matching is not fulfilled [Fig.~\ref{peakspump} (b)]. At 
present, we cannot explain these noise features. Further investigation is needed, in order to reveal the 
physical mechanism behind such rich structure.

\subsection{New quantum correlations} 

\begin{figure}[ht]
\centering
\epsfig{file=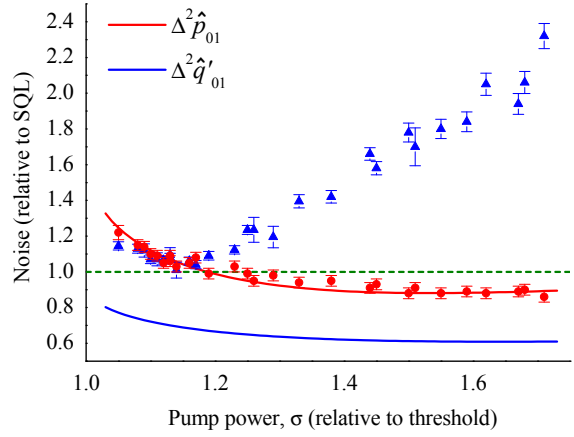,scale=0.45}
\caption{Behaviors of the noise terms appearing in Ineq~(\ref{crit2}) as functions of pump power 
$\sigma$. The red circles correspond to the first term in the inequality and the blue triangles 
to the second one. The former presents the observation of amplitude quantum correlations 
between one of the twin beams and the reflected pump beam: there is $\approx$ 12\% of squeezing 
for higher $\sigma$ values. The solid lines are the predictions of the model explained in the 
text, without added noise. Theoretical curves have the same color as the experimental symbol 
to which they relate.}
\label{behavsigmapump}
\end{figure}

In spite of these unknown noise features, we could still check our experimental results against 
the tripartite criteria of Ineqs.~(\ref{crit1}) -- (\ref{crit3}). We have already shown violation 
of Ineq.~(\ref{crit1}). Owing to the similarity between the twin beams, it suffices to investigate 
Ineq.~(\ref{crit2}). Our experimental results, as a function of $\sigma$, are presented in 
Fig.~\ref{behavsigmapump}. A new quantum correlation, among the amplitudes of the pump and the 
signal beams (manifest in $\Delta^2\p_{01}$, red circles), is observed for the first time in the 
above-threshold OPO. We obtain squeezing for higher values of pump power, $\Delta^2\p_{01}\approx0.88(2)$. 
Since it only involves amplitude quadratures, good agreement with the 
theoretical model (full red line) is obtained (no excess noise is input in the model). The sum 
of this term with $\Delta^2\q'_{01}$ (phase difference between pump and signal phases, corrected 
by idler phase noise) corresponds to the left hand side in Ineq.~(\ref{crit2}). Data for 
$\Delta^2\q'_{01}$ are presented as blue triangles in Fig.~\ref{behavsigmapump}. The solid blue 
line corresponds to the theoretical model with no extra noise (adding extra noise does not provide 
good agreement). For $\sigma \approx 1.15$, we approach the classical limit of Ineq.~(\ref{crit2}), 
with the lowest value given by $V_1$=2.04(10). 

Even by including spurious excess noise in the model, we would expect lower 
noise values for $\Delta^2\q'_{01}$, leading to the predicted tripartite entanglement. Although 
we observed two new quantum correlations among the pump and the twin beams, genuine tripartite 
entanglement still remains to be demonstrated. Our results suggest that minor modifications 
to the setup may suffice. Technical improvements, such as a reduction of intracavity spurious 
losses and of detection losses, entail better squeezing and better detection efficiency. 

The ability to measure any given quadrature of each field and their combinations, together 
with the broad range of pump powers relative to threshold that we can access, allow for complete 
characterizations of the OPO's noise properties. In the near future, we expect to determine 
the excess noise source, to circumvent it, and reliably generate tripartite entanglement.

\section{Conclusion}

We presented a new and thorough investigation of quadrature noise of pump, signal, and idler 
fields in the above-threshold OPO. Although the OPO theory is well established, an 
important part of it had never been tested above the oscillation threshold. We revealed 
an unnoticed extra noise in the system and have further characterized several of its 
features. It seems safe to say that only phase noise is generated in the intracavity 
crystal. The amplitude noise observed in previous experiments was due to phase-to-amplitude 
conversion by the quasi-resonant cavity. This effective refractive index should be frequency dependent. 
Thus, by changing the pump frequency, it may be possible to minimize it. All this is still 
insufficient to pin down the ultimate physical origin of this noise. More research is needed 
to solve this intriguing puzzle. 

Notwithstanding the unwelcome extra noise, we could demonstrate new quantum correlations 
among pump and twin beams. The predicted quantum correlation among the phase quadratures, 
which is a direct consequence of energy conservation, was observed for the first time. 
When operating further above threshold, pump depletion becomes more evident as we observe the 
quantum correlation between the pump and each twin's amplitudes. By combining these results, 
we approached the limit for demonstrating tripartite entanglement: $V_0$=1.29(4), $V_1$=2.04(10), 
and $V_2$=2.09(7), where a violation corresponds to $V_j <2 $. Hopefully, by improving the 
squeezing and the detection efficiencies, we will beat this limit. 

The approach to directly generate multipartite entanglement, when compared to its 
generation by means of individual squeezers and linear optics, has the potential 
advantage of entangling fields of different colors. This advantage is exploited in 
our setup and we expect it to be the starting point for multicolor quantum networks.

\subsection*{Acknowledgments} 
This work was supported by Funda\c{c}\~ao de Amparo \`a Pesquisa do Estado de S\~ao Paulo 
(FAPESP), Conselho Nacional de Desenvolvimento Cient\'\i fico e Tecnol\'ogico (CNPq), through 
{\it Instituto do Mil\^enio de Informa\c c\~ao Qu\^antica}, and Coordena\c{c}\~ao de 
Aperfei\c{c}oamento de Pessoal de N\'{\i}vel Superior (CAPES). 


\begin{thebibliography}{99}
\bibitem{firsttwin87}
A. Heidmann, R. J. Horowicz, S. Reynaud, E. Giacobino, C. Fabre, and G. Camy,
``Observation of Quantum Noise Reduction on Twin Laser Beams'',
Phys. Rev. Lett. {\bf 59,} 2555--2557 (1987).
 
\bibitem{fabre10dB} J. Laurat, L. Longchambon, C. Fabre, and T. Coudreau, 
``Experimental investigation of amplitude and phase quantum correlations in a type II optical
parametric oscillator above threshold: from nondegenerate to degenerate operation'',
Opt. Lett. {\bf 30,} 1177--1179 (2005).
 
\bibitem{reiddrummondepropo_prl88}
M. D. Reid  and P. D. Drummond,
``Quantum correlations of phase in nondegenerate parametric oscillation'',
Phys. Rev. Lett. {\bf 60,} 2731--2733 (1988)

\bibitem{kimbleepr} Z. Y. Ou, S. F. Pereira, H. J. Kimble, and K. C. Peng, 
``Realization of the Einstein-Podolsky-Rosen Paradox for Continuous Variables'', 
\prl \textbf{68,} 3663--3666 (1992).
 
\bibitem{furusawatele_science98}
A. Furusawa, J. L. S\o{}rensen, S. L. Braunstein, C. A. Fuchs, H. J. Kimble, and E. S. Polzik,
``Unconditional Quantum Teleportation'',
Science {\bf 282,} 706--709 (1998).

\bibitem{pengdensecoding_prl02}
X. Li, Q. Pan, J. Jing, J. Zhang, C. Xie, and K. Peng,
``Quantum Dense Coding Exploiting a Bright Einstein-Podolsky-Rosen Beam'',
Phys. Rev. Lett. {\bf 88,} 047904 (2002).

\bibitem{furusawanetwork_nature04}
H. Yonezawa, T. Aoki, and A. Furusawa,
``Demonstration of a quantum teleportation network for continuous variables'',
Nature {\bf 431,} 430--433 (2004).

\bibitem{prlentangtwinopo} 
A. S. Villar, L. S. Cruz, K. N. Cassemiro, M. Martinelli, and P. Nussenzveig, 
``Generation of Bright Two-Color Continuous Variable Entanglement'', 
Phys. Rev. Lett. {\bf 95,} 243603 (2005). 

\bibitem{optlettpeng} 
X. L. Su, A. Tan, X. J. Jia, Q. Pan, C. D. Xie, and K. C. Peng, 
``Experimental demonstration of quantum entanglement between frequency-nondegenerate optical twin beams'', 
Opt. Lett. \textbf{31,} 1133--1135 (2006).

\bibitem{pfisterentang} 
J. Jing, S. Feng, R. Bloomer, and O. Pfister, 
``Experimental continuous-variable entanglement of phase-locked bright optical beams'', 
Phys. Rev. A {\bf 74,} 041804(R) (2006).

\bibitem{josaboptquinfo} 
A. S. Villar, K. N. Cassemiro, K. Dechoum, A. Z. Khoury, M. Martinelli, and P. Nussenzveig, 
``Entanglement in the above-threshold optical parametric oscillator'',
J. Opt. Soc. Am. B {\bf 24,} 249--256 (2007).

\bibitem{kasaisqzpump_europhyslett97}
K. Kasai, G. Jiangrui, and C. Fabre, 
``Observation of squeezing using cascaded nonlinearity''
Europhys. Lett. {\bf 40,} 25--30 (1997).

\bibitem{prltrientangopo} 
A. S. Villar, M. Martinelli, C. Fabre, and P. Nussenzveig, 
``Direct Production of Tripartite Pump-Signal-Idler Entanglement in the Above-Threshold OPO'', 
Phys. Rev. Lett. {\bf 97,} 140504 (2006). 

\bibitem{optletttripqucorr}
K. N. Cassemiro, A. S. Villar, P. Valente, M. Martinelli, and P. Nussenzveig, 
``Experimental observation of three-color optical quantum correlations'',
Opt. Lett. {\bf 32,} 695--697 (2007).

\bibitem{vanloockmultipartite_pra03}
P. van Loock and A. Furusawa,
``Detecting genuine multipartite continuous-variable entanglement'',
Phys. Rev. A {\bf 67,} 052315 (2003).

\bibitem{dgcz_prl00} 
L. M. Duan, G. Giedke, J. I. Cirac, and P. Zoller,
``Inseparability Criterion for Continuous Variable Systems'', 
Phys. Rev. Lett. {\bf 84,} 2722--2725 (2000).

\bibitem{simon_prl00} 
R. Simon, 
``Peres-Horodecki Separability Criterion for Continuous Variable Systems'', 
Phys. Rev. Lett. {\bf 84,} 2726--2729 (2000).

\bibitem{modecleancavity}
B. Willke, N. Uehara, E. K. Gustafson, R. L. Byer, P. J. King, S. U. Seel, and R. L. Savage,
``Spatial and temporal filtering of a 10-W Nd:YAG laser with a Fabry-Perot ring-cavity premode cleaner'',
Opt. Lett. {\bf 23,} 1704--1706 (1998).

\bibitem{optcomm04} 
A.S. Villar, M. Martinelli, and P. Nussenzveig,
``Testing the entanglement of intense beams produced by a
non-degenerate optical parametric oscillator'',
Opt. Comm. {\bf 242,} 551–-563 (2004).

\bibitem{galatola_optcomm91}
P. Galatola, L. A. Lugiato, M. G. Porreca, P. Tombesi, and G. Leuchs,
``System control by variation of the squeezing phase'',
Opt. Commun. {\bf 85,} 95--103 (1991).

\end{thebibliography}
\end{document}